\documentclass[showpacs,aps,graphicx,twocolumn]{revtex4}%,,pdflatex
\usepackage{amsmath}%,preprint,twocolumn,
\usepackage{graphicx}

\begin{document}

%Robust assisted by  quantum-dot spins in optical microcavities

\title{Systematic entanglement concentration for unknown less-entangled three-photon W states\footnote{Published in Laser Phys. Lett. \textbf{12}, 115202 (2015).}}

\author{Fang-Fang Du$^1$ and Fu-Guo Deng$^{1,2,}$\footnote{Corresponding author: fgdeng@bnu.edu.cn} }

\address{$^1$Department of Physics, Applied Optics Beijing Area Major Laboratory,
Beijing Normal University, Beijing 100875, China\\
$^2$State key Laboratory of Networking and Switching Technology,
Beijing University of Posts and Telecommunications, Beijing 100876,
China}

\date{\today }

%in the first step of the first round of concentration

%. These advantages maybe make our ECP more feasible in current
%quantum information processing. It

\begin{abstract}
We present a systematic entanglement concentration protocol  (ECP)
for an arbitrary unknown less-entangled three-photon W state,
resorting to the optical property of the quantum-dot spins inside
one-sided optical microcavities. In our ECP, the parties obtain not
only some three-photon systems in the partially entangled with two
unknown parameters when one of the parties picks up the robust
odd-parity instance with the parity-check gate (PCG) on his two
photons, but also some entangled two-photon systems by keeping the
even-parity instance in the first step. By exploiting the above
three-photon and two-photon systems with the same parameters as the
resource for the second step of our ECP, the parties can obtain a
standard three-photon W state by keeping the robust odd-parity
instance. Meanwhile, the systems in the even-parity instance can be
used as the resource in the next round of our ECP. The success
probability of our ECP  is  largely increased by iteration of the
ECP process. As it does require that  all  the coefficients are
unknown for the parties, our ECP maybe have good applications in
quantum communication network in future.
\end{abstract}

\maketitle

%
%
%
%\textbf{Keywords} Entanglement concentration. Quantum communication
%. Quantum dot. One-sided optical microcavity. Three-photon systems
%.Quantum electrodynamics

\section{Introduction}

Quantum communication becomes one of the two important branches of
quantum information and it has attracted much attention, especially
teleportation  \cite{c1}, dense coding  \cite{dense1,dense2},
quantum key distribution (QKD)  \cite{BB84,QKD1,QKD2,QKD3}, quantum
secret sharing  \cite{QSS1}, and quantum secure direct communication
 \cite{QSDC1,QSDC2,QSDC3}. Quantum entanglement is a key
resource for quantum communication. It can act as the information
carrier in some important quantum communication protocols
 \cite{QKD1,QKD2,QKD3,QSS1,QSDC1,QSDC2}. In practical
long-distance quantum communication, quantum repeaters are required
to overcome the photon loss and decoherence from environment noise,
in which entanglement is used to create the quantum channel between
two neighboring quantum nodes. That is to say, all the tasks in
long-distance quantum communication should require high-fidelity
entanglement. However, entanglement can only be produced locally and
it is very fragile in the process of transmission and storage, due
to the influence of decoherence and the imperfection at the source.
The decoherence of  entangled quantum systems will decrease the
security of QKD protocols and even makes a quantum teleportation and
a quantum dense coding protocol fail.

There are some useful methods to depress the above fragile effect on
entangled photon systems, such as conventional entanglement
purification \cite{Bennett,panepp,simonepp,shengpra} with which the
remote users can obtain some high-fidelity entangled systems from
low-fidelity ones, deterministic entanglement purification
\cite{shengpratwostep,shengpraonestep,lixhonestep,dengonestep,DEPPShengLPL,DEPPShengSR}
(which works  in a completely deterministic way, not in a
probabilistic way, and they can reduce the quantum resource
sacrificed largely \cite{gooddepp}), and entanglement concentration
\cite{BennettECP,YamamotoECP,zhaoECP,shengpraEC1,Bose,swappECP,shengpraEC3,dengECP,HECPsplitting,Heppren1,renoeecp,hyperconcentration1,hperECPlixh2,
hperECPlixh3}. By definition, entanglement purification protocols
are used to distill some high-fidelity entangled photon systems from
a less-entangled ensemble in a mixed entangled state
 \cite{Bennett,panepp,simonepp,shengpra,shengpratwostep,shengpraonestep,lixhonestep,dengonestep,DEPPShengLPL,DEPPShengSR}
while entanglement concentration protocols (ECPs)
 \cite{BennettECP,YamamotoECP,zhaoECP,shengpraEC1,Bose,swappECP,shengpraEC3,dengECP,HECPsplitting,Heppren1,renoeecp,hyperconcentration1,hperECPlixh2,
hperECPlixh3,CaoCatomecpOE,wangECP,wangECPPRA,electronecp,ecpab2,ecpa11}
are used to obtain a subset of photon systems in a maximally
entangled state from a set of systems in a partially entangled pure
state. The former is more general as a photon system is usually in a
mixed entangled state after it is transmitted over a noisy channel,
but the latter  is a more efficient in some particular cases, such
as those with decoherence of entanglement arising from the storage
process or the imperfect entanglement source.

The first ECP was proposed by Bennett \emph{et al} \cite{BennettECP}
in 1996. In 2001, two ECPs based on polarizing beam splitters were
proposed \cite{YamamotoECP,zhaoECP}. In 2008, Sheng, Deng, and Zhou
\cite{shengpraEC1} proposed a repeatable ECP  which has a far higher
efficiency and yield than the PBS-ECPs \cite{YamamotoECP,zhaoECP},
by iteration of the entanglement concentration process three times.
In fact, depending on whether the parameters of the less-entangled
states are known
\cite{Bose,swappECP,shengpraEC3,dengECP,HECPsplitting} or not
\cite{BennettECP,YamamotoECP,zhaoECP,shengpraEC1}, the ECPs can be
classed into two groups. When the parameters are known, one nonlocal
photon system is enough for entanglement concentration
\cite{Bose,swappECP,shengpraEC3,dengECP,HECPsplitting}, far more
efficient that those with unknown parameters
\cite{BennettECP,zhaoECP,YamamotoECP,shengpraEC1}.   In 2013, Ren,
Du, and Deng \cite{HECPsplitting} proposed the parameter-splitting
method to extract the maximally entangled photons in both the
polarization and spatial degrees of freedom (DOFs) when the
coefficients of the initial partially hyperentangled states are
known. This fascinating method is very efficient and simple in terms
of concentrating partially entangled states, and it can be achieved
with the maximum success probability by performing the protocol only
once, resorting to linear optical elements only, not nonlinearity
\cite{goodecp}. They \cite{HECPsplitting} also gave the first
hyperentanglement concentration protocol (hyper-ECP) for the known
and unknown polarization-spatial less-hyperentangled states with
linear-optical elements only. Recently, some good hyper-ECPs
\cite{Heppren1,renoeecp,hyperconcentration1,hperECPlixh2,hperECPlixh3}
for photon systems were proposed.  Now, some efficient  ECPs for
atomic systems \cite{CaoCatomecpOE} and electronic systems
\cite{wangECP,wangECPPRA,ecpab2,electronecp}  have been proposed.

By far, there  are few ECPs for entangled pure W-class states
 \cite{Yildiz,shengpraEC4,wang2,ecpaa6,du,ecpgujosab,ecpaa10,ecpaa8,ecpxiayqip}.
In essence, W states are inequivalent to Greenberger-Horne-Zeilinger
(GHZ) states, as they cannot be converted into each other under
stochastic local operations and classical communication (LOCC).
Moreover, tripartite W entanglement has both bipartite and
tripartite quantum entanglement simultaneously, thus it is robust to
the loss of one qubit\cite{Vidal} and can be used in some quantum
information processing such as unconditionally teleclone coherent
states  \cite{teleclone}, probabilistic teleportation of unknown
atomic state\cite{teleportation}, quantum information sharing
\cite{AugusiakSKD}, random-party
distillation\cite{distillationL,distillationA}, and so on. That is,
it is of practical significance to discuss the entanglement
concentration on the partially entangled W state.  In 2010, Yildiz
 \cite{Yildiz} proposed an optimal distillation of three-qubit
asymmetric W states. Based on linear optical elements, an ECP for
partially entangled W states are proposed by Wang \emph{et al}
\cite{wang2}. Subsequently, Du \emph{et al} \cite{du} and Gu
\emph{et al}
 \cite{ecpaa6} improved the ECP for the special W states by
exploiting the cross-Kerr nonlinearity, respectively.  However,
these ECPs  \cite{wang2,du,ecpaa6} are used to deal with the
concentration on three-photon systems in a partially entangled W
state with only two parameters, not arbitrary coefficients. In 2012,
Sheng \emph{et al} \cite{shengpraEC4} proposed an ECP for
three-photon systems in an arbitrary less-entangled W-type state
with the known coefficients by assisting two different specially
polarized single-photon sources. Based on linear optical elements,
Wang and Long \cite{ecpaa8} presented two three-photon ECPs for an
arbitrary unknown less-entangled W-class state in 2013.

Recently, a single spin coupled to an optical microcavity based on a
charged self-assembled GaAs/InAs quantum dot (QD) has attracted much
attention as it is a novel candidate for a quantum qubit. Since Hu
\emph{et al} \cite{HuQD1} pointed out that the interaction between a
circularly polarized light and a QD-cavity system can be used for
quantum information processing,  some ECPs
\cite{wangECP,wangECPPRA,renoeecp,ecpaa10} have been proposed with
this system. For example, in 2011, Wang \emph{et al}
\cite{wangECPPRA} proposed an ECP based on QD-microcavity systems
with two copies of partially entangled two-electron systems to
obtain a maximally entangled two-electron system probabilistically.
Subsequently, Wang \cite{wangECP} showed that each two-electron-spin
system in a partially entangled state can be concentrated with the
assistance of an ancillary quantum dot and a single photon, not two
copies of two-electron spin systems. In 2013, Sheng \emph{et al}
\cite{ecpaa10} proposed an efficient ECP for  W-class states
assisted by  the double-sided optical microcavities.

In this paper, we propose a systematic ECP for an arbitrary unknown
less-entangled three-photon W state, resorting to the optical
property of the quantum-dot spins inside one-sided optical
microcavities. The parties obtain not only some partially entangled
three-photon systems with two unknown parameters when one of the
parties picks up the robust odd-parity instance with the
parity-check gate (PCG) on his two photons, but also some entangled
two-photon systems by keeping an even-parity instance in the first
step of the first round of concentration. By exploiting the above
three-photon and two-photon systems with the same parameters as the
resource for the second step of our ECP, the parties can obtain a
standard three-photon W state by keeping the robust odd-parity
instance, far different to the previous ECPs for W-class states with
unknown parameters \cite{Yildiz}.  Meanwhile, the systems in the
even-parity instance can be used as the resource in the next round
of the ECP.  The success probability of our ECP is largely increased
by iteration of the ECP process. Besides, as the side leakage and
cavity loss may be difficult to control or reduce for the  photonic
qubits in the double-sided QD-cavity system, our ECP is relatively
easier to be implemented in experiment than the ECP with a
double-sided QD-cavity system \cite{ecpaa10}. These advantages maybe
make our ECPs more useful in quantum communication network in
future.

\section{Parity-check gate on two photons assisted by a QD-cavity system}

\subsection{Interaction between a circularly polarized light and a
QD-cavity system}  \label{sec2}

\begin{figure}[tpb]           %Figure  1 for the exciton energy levels
\begin{center}
\includegraphics[width=5.8 cm,angle=0]{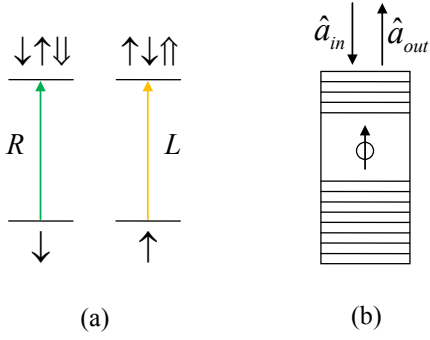}
\caption{The spin-dependent transitions for negatively charged
exciton $X^{-}$. (a) Spin selection rule for optical transitions of
negatively charged exciton $X^{-}$ due to the Pauli's exclusion
principle. (b) A charged QD inside a micropillar microcavity with
circular cross section. L and R represent the left and the right
circularly polarized lights, respectively. $\uparrow$ and
$\downarrow$ represent the spins of the excess electron.
$\downarrow\uparrow\Downarrow$ and $\uparrow\downarrow\Uparrow$
represent the negatively charged exciton $X^{-}$. } \label{fig1}
\end{center}
\end{figure}

The solid-state system discussed here is a singly charged QD in a
one-sided  cavity, e.g.,  a self-assembled In(Ga)As QD or a GaAs
interface QD located in the center of an optical resonant cavity
(the bottom distributed Bragg reflectors are $100\%$ reflective and
the top distributed Bragg reflectors are partially reflective) to
achieve a maximal light-matter coupling \cite{HuQD1}, shown in Fig.
\ref{fig1}(b). If an excess electron is injected into the QD, the
optical resonance can create the negatively charged exciton $X^{-}$
that consists of two electrons bound to one hole \cite{Warburton}.
This means that $X^{-}$ has the spin-dependent optical transitions
\cite{Hu} (shown in Fig. \ref{fig1}(a)) for the circularly polarized
probe lights, according to Pauli's exclusion principle. The
left-circularly polarized light $|L\rangle$ is resonantly absorbed
to create the negatively charged exciton
$|\uparrow\downarrow\Uparrow\rangle$ for the excess electron spin
state $|\uparrow\rangle_e$, and the right-circularly polarized light
$|R\rangle$ is resonantly absorbed to create the negatively charged
exciton $|\downarrow\uparrow\Downarrow\rangle$ for the excess
electron spin state $|\downarrow\rangle_e$. Here $|\Uparrow\rangle$
($|\Downarrow\rangle$) represents the heavy-hole spin state
$|+\frac{3}{2}\rangle$ ($|-\frac{3}{2}\rangle$). This process can be
described by the Heisenberg equations for the cavity field operator
$a$ and $X^{-}$ dipole operator $\sigma_{-}$ in the interaction
picture ($\hbar=1$) \cite{Walls-QD},
\begin{eqnarray}                     \label{H}   % Eq. 1
\begin{split}
\frac{da}{dt}&=-\left[i(\omega_{c}-\omega)+\frac{\kappa}{2}+\frac{\kappa_{s}}{2}\right]a-g\,\sigma_{-}-\sqrt{\kappa}\,a_{in},\\
\frac{d\sigma_{-}}{dt}&=-\left[i(\omega_{X^{-}}-\omega)+\frac{\gamma}{2}\right]\sigma_{-}-g\sigma_{z}\,a,\\
a_{out}&=a_{in}+\sqrt{\kappa}\,a.
\end{split}
\end{eqnarray}
Here, $\omega$, $\omega_{X^{-}}$, and $\omega_{c}$ are the
frequencies of the input probe light, $X^{-}$ transition, and cavity
mode, respectively. $\frac{\gamma}{2}$ and $\frac{\kappa}{2}$ are
the decay rates of $X^{-}$ and the cavity field, respectively.
$\frac{\kappa_{s}}{2}$ is the side leakage rate of the cavity.  $g$
is the coupling strength between $X^{-}$ and the cavity mode.

Considering a weak excitation condition with $X^{-}$ staying in the
ground state at most time and $\langle\sigma_{z}\rangle=1$, the
reflection coefficient of circularly polarized light after
interacting with a QD-cavity system is \cite{HuQD1,Walls-QD}
\begin{eqnarray}                     \label{rh} % Eq. 2
r_{h}(\omega)=1\!-\!\frac{\kappa\left[i(\omega_{X^{-}}-\omega)
+\frac{\gamma}{2}\right]}{\left[i(\omega_{X^{-}}\!-\!\omega)\!+\!\frac{\gamma}{2}\right]
\left[i(\omega_{c}\!-\!\omega)\!+\!\frac{\kappa}{2}\!+\!\frac{\kappa_{s}}{2}\right]
\!+\!g^{2}}.\nonumber\\
\end{eqnarray}
If the $X^{-}$ is uncoupled to the cavity, which is called a cold
cavity with the coupling strength $g=0$, the reflection coefficient
becomes \cite{HuQD1,Walls-QD}
\begin{eqnarray}         \label{r0} % Eq. 3
r_{0}(\omega)=\frac{i(\omega_{c}-\omega)-\frac{\kappa}{2}
+\frac{\kappa_{s}}{2}}
{i(\omega_{c}-\omega)+\frac{\kappa}{2}+\frac{\kappa_{s}}{2}}.
\end{eqnarray}
The polarized light may have a phase shift after being reflected
from the QD-cavity system. By adjusting the frequencies $\omega$ and
$\omega_{c}$ and neglecting the cavity side leakage, one can get
$|r_{0}(\omega)|\cong 1$ for a cold cavity and $|r_{h}(\omega)|\cong
1$ for a hot cavity. The $|L\rangle$ light gets a phase shift of
$\varphi_{h}$ for a hot cavity when the excess electron spin state
is $|\uparrow\rangle_e$, and it gets a phase shift of $\varphi_{0}$
for a cold cavity when the excess electron spin state is
$|\downarrow\rangle_e$. Conversely, the $|R\rangle$ light gets a
phase shift of $\varphi_{0}$ for a cold cavity when the excess
electron spin state is $|\uparrow\rangle_e$, and it gets a phase
shift of $\varphi_{h}$ for a hot cavity when the excess electron
spin state is $|\downarrow\rangle_e$. Therefore, the superposition
of two circularly polarized probe beams
$(|R\rangle+|L\rangle)/\sqrt{2}$ becomes
$(e^{i\varphi_{0}}|R\rangle+e^{i\varphi_{h}}|L\rangle)/\sqrt{2}$ for
the electron spin state $|\uparrow\rangle_e$ and
$(e^{i\varphi_{h}}|R\rangle+e^{i\varphi_{0}}|L\rangle)/\sqrt{2}$ for
the electron spin state $|\downarrow\rangle_e$ after being reflected
from the QD-cavity system. The Faraday rotation is defined by the
rotation angle of the polarization direction
$\theta_{F}^{\uparrow}=(\varphi_{0}-\varphi_{h})/2=\theta_{F}^{\downarrow}$.
If a polarized probe beam $(|R\rangle+|L\rangle)/\sqrt{2}$ is put
into the QD-cavity system with the electron spin in the state
$(\alpha|\uparrow\rangle+\beta|\downarrow\rangle)_e$, after
reflection, the state of the system composed of the light and the
electron spin becomes entangled,
\begin{eqnarray}          \label{R}   % Eq. 4
&&(|R\rangle+|L\rangle)/\sqrt{2}\otimes(\alpha|\uparrow\rangle+\beta|\downarrow\rangle)_e
\;\;\longrightarrow\;\; \nonumber\\
&&e^{i\varphi_{0}}\!\! \left[\alpha(|R\rangle\!+\!e^{i
\Delta\varphi}|L\rangle)|\uparrow\rangle_e \! +\! \beta(e^{i
\Delta\varphi}|R\rangle\!+\!|L\rangle)|\downarrow\rangle_e\right]/\sqrt{2}.\nonumber\\
\end{eqnarray}
Here, $\Delta\varphi=\varphi_{h}-\varphi_{0}$,
$\varphi_{0}=arg[r_{0}(\omega)]$ and $\varphi=arg[r(\omega)]$. Due
to the spin-selection rule, $|R\rangle$ and $|L\rangle$ lights
encounter different phase shifts after being reflected from the
one-sided QD-cavity system, and the state of the system composed of
the light and the spin in QD becomes entangled.

If the frequencies of the input light and cavity mode are adjusted
as $\omega-\omega_{c}\approx\kappa/2$, the relative phase shift of
the left and right circularly polarized lights is
$\Delta\phi=\frac{\pi}{2}$. The rules of the input photon states
changing under the interaction of the photon and the QD-cavity
system can be described as follows:
\begin{eqnarray}          \label{RULE}   % Eq. 5
\begin{split}
& |L,\uparrow\rangle\rightarrow |L,\uparrow\rangle, \;\;\;\;\;\;
|L,\downarrow\rangle\rightarrow i|L,\downarrow\rangle,
\;\;\;\; \\
& |R,\uparrow\rangle\rightarrow i|R,\uparrow\rangle, \;\;\;\;
|R,\downarrow\rangle\rightarrow |R,\downarrow\rangle.
\end{split}
\end{eqnarray}

\subsection{Parity-check gate on the polarizations of two photons}

\begin{figure}[tpb]           %Figure 2 for PCG
\begin{center}
\includegraphics[width=8 cm,angle=0]{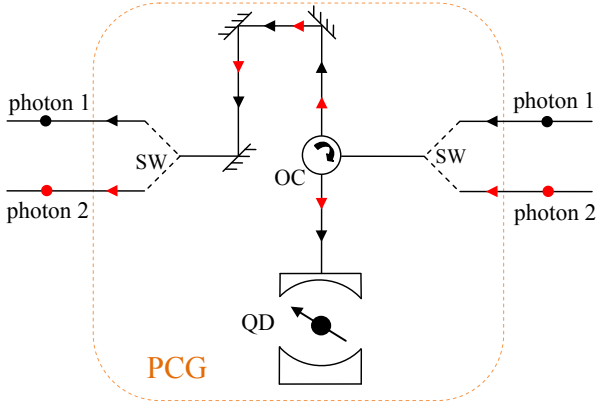}
\caption{ Schematic diagram of the parity-check gate on the
polarizations of two photons, assisted by  a QD-spin in a
single-sided microcavity  with  the top mirror partially reflective
and the bottom mirror $100\%$ reflective. The optical switch (SW)
and the optical circulator (OC) make photon 1 and photon 2 pass
through the cavity in sequence.} \label{fig2}
\end{center}
\end{figure}

The principle of our parity-check gate on the polarizations of two
photons is shown in  Fig.\ref{fig2}. Suppose that the electron spin
in the QD is prepared in the superposition state
$|\psi\rangle^{s}_e=\frac{1}{\sqrt{2}}(|\uparrow\rangle+|\downarrow\rangle)_e$.
Photon 1 in the state
$|\psi\rangle^{p}_{1}=\alpha_{1}|R\rangle_{1}+\beta_{1}|L\rangle_{1}$
and photon 2 in the state
$|\psi\rangle^{p}_{2}=\frac{1}{\sqrt{2}}(|R\rangle_{2}+|L\rangle_{2})$
enter into the QD-cavity system in sequence. An optical circulator
(OC) first directs  photon 1 until it is reflected by the cavity,
and then it is switched for photon 2. The time difference between
photons 1 and 2 should be less than the spin coherence time. After
two photons are reflected, the state of the composite system
composed of two photons and one QD-spin is evolved as follows:
\begin{eqnarray}   \label{PCG1}    % Eq. 6
\begin{split}
|S\rangle_0 \equiv&\; |\psi\rangle^{p}_{1}
\otimes|\psi\rangle^{p}_{2}\otimes|\psi\rangle^{s}_e
 \;\;\rightarrow\;\;\\
|S\rangle_1 = &\;
\frac{1}{2}(|\uparrow\rangle-|\downarrow\rangle)_e
(\alpha_{1}|R\rangle_{1}|R\rangle_{2}-\beta_{1}|L\rangle_{1}|L\rangle_{2})
\\
&+\frac{i}{2}(|\uparrow\rangle+|\downarrow\rangle)_e
(\alpha_{1}|R\rangle_{1}|L\rangle_{2}+\beta_{1}|L\rangle_{1}|R\rangle_{2}).%\nonumber\\
\end{split}
\end{eqnarray}
By detecting the electron-spin state, one can distinguish the
even-parity state $\vert \phi_{12}
\rangle_{even}=\alpha_{1}|R\rangle_{1}|R\rangle_{2}-\beta_{1}|L\rangle_{1}|L\rangle_{2}$
of the two-photon system from the odd-parity state $\vert
\psi_{12}\rangle_{odd}=\alpha_{1}|R\rangle_{1}|L\rangle_{2}+\beta_{1}|L\rangle_{1}|R\rangle_{2}$.
This task can be achieved with a probe photon 3 which interacts with
the electron spin (the GFR-based quantum nondemolition method)
\cite{HuQD3}. In detail, suppose that the photon 3 is initially in
the state
$|\psi\rangle^{p}_{3}=\frac{1}{\sqrt{2}}(|R\rangle_{3}+|L\rangle_{3})$.
One performs a Hadamard transformation [$|\uparrow\rangle_e
\rightarrow |+\rangle_e=\frac{1}{\sqrt{2}}
(|\uparrow\rangle+|\downarrow\rangle)_e$, $|\downarrow\rangle_e
\rightarrow |-\rangle_e=\frac{1}{\sqrt{2}}
(|\uparrow\rangle-|\downarrow\rangle)_e$, e.g.,  using a $\pi/2$
microwave pulse] on the electron spin before photon 3 is input into
the cavity (the photon 3 has the same frequency as the photons 1 and
2).  After the photon 3 is reflected, the state of the composite
system composed of the three photons and one electron spin becomes
\begin{eqnarray}     \label{PCG11}     % Eq. 7
|S\rangle_2\!\!&=&\!\!\frac{1}{2}(|L\rangle_{3}+i|R\rangle_{3})|\downarrow\rangle_e
(\alpha_{1}|R\rangle_{1}|R\rangle_{2}-\beta_{1}|L\rangle_{1}|L\rangle_{2})
\nonumber\\
&&-\frac{1}{2}(|L\rangle_{3}-i|R\rangle_{3})|\uparrow\rangle_e
(\alpha_{1}|R\rangle_{1}|L\rangle_{2}+\beta_{1}|L\rangle_{1}|R\rangle_{2}).\nonumber\\
\end{eqnarray}
The output state of photon 3 can be measured in orthogonal linear
polarizations.  If  photon  3 is detected in the
$(|L\rangle_{3}+i|R\rangle_{3})/\sqrt{2}$ state ($45^{\circ}$
linear), the electron spin is  in the state $|\downarrow\rangle_e$
and the system composed of photon  1 and photon  2 is in the
even-parity  state $\vert \phi_{12}
\rangle_{even}=\alpha_1|R\rangle_1|R\rangle_2 -
\beta_1|L\rangle_1|L\rangle_2$. Otherwise, if photon 3 is detected
in the $(|L\rangle_{3}-i|R\rangle_{3})/\sqrt{2}$ state
($-45^{\circ}$ linear),  the electron spin is in the state
$|\uparrow\rangle_e$ and the two-photon system is in the odd-parity
state $\vert \psi_{12}\rangle_{odd}=\alpha_1|R\rangle_1|L\rangle_2 -
\beta_1|R\rangle_1|L\rangle_2$.

\begin{figure}[htp]
\begin{center}
\includegraphics[width=7 cm,angle=0]{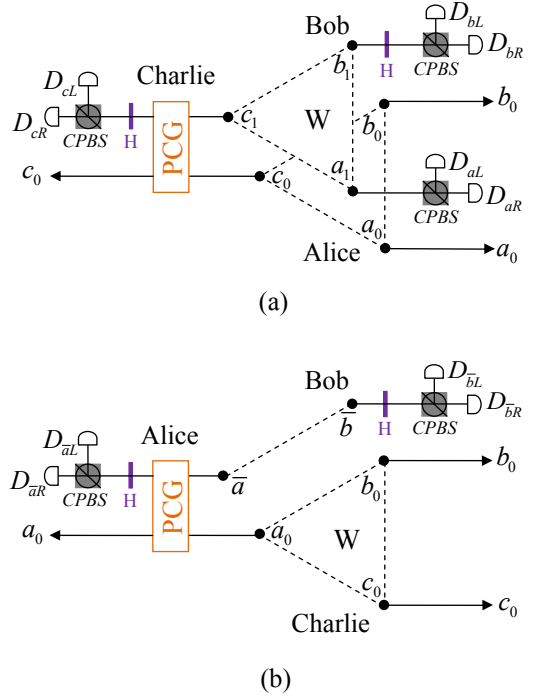}
\caption{ Schematic diagram of  our ECP for three-photon systems in
a less-entanglement W state with the parity-check gate (PCG) on the
polarizations of two photons. (a) The first step of our ECP. (b) The
second step of our ECP. CPBS represents a polarizing beam splitter
in the circularly polarized basis $\{|R\rangle,|L\rangle\}$, which
transmits the $|R\rangle$ polarization photons and reflects the $|L
\rangle$ polarization photons. H represents a Hadamard operation on
the polarization of the photon. $D_{mL}$ and $D_{mR}$, $(m\in \{a,
b, c, \overline{a}, \overline{b}\})$ are  single-photon detectors. }
\label{fig3}
\end{center}
\end{figure}

%\subsection*{3.2 $\;\;$  ECP for less-entangled three-photon W state} \label{sec3.2}

\section{Systematic ECP for less-entangled three-photon W state}

Let us assume that the three-photon system $a_{i}b_{i}c_{i}$
($i=0,1,2, \cdots$) is in the following less-entangled state:
\begin{eqnarray}         \label{eq1}    % Eq. 8
|\Phi\rangle_{a_{i}b_{i}c_{i}}&=&\alpha|R\rangle_{a_{i}}|R\rangle_{b_{i}}|L\rangle_{c_{i}}
+\beta|R\rangle_{a_{i}}|L\rangle_{b_{i}}|R\rangle_{c_{i}}
\nonumber\\
&&+\gamma|L\rangle_{a_{i}}|R\rangle_{b_{i}}|R\rangle_{c_{i}},
\end{eqnarray}
where the subscripts $a_{i}$, $b_{i}$, and $c_{i}$ represent the
three photons belonging to the three remote parties in quantum
communication, say Alice, Bob, and Charlie, respectively. Here
$\alpha$, $\beta$, and $\gamma$ are three arbitrary unknown
parameters and   they  satisfy the relation $|\alpha|^{2}+
|\beta|^{2}+|\gamma|^{2}=1$. The principle of our systematic ECP
with nonlinear optical elements is shown in Fig.\ref{fig3}, and it
includes two steps. We describe it in detail as follows.

The principle for the first step of our systematic ECP is shown in
Fig.\ref{fig3} (a).  In this step, in each round of entanglement
concentration,  Alice, Bob and Charlie operate two pairs of
three-photon less-entangled systems, say  $a_{0}b_{0}c_{0}$ and
$a_{1}b_{1}c_{1}$. The state
$|\Phi\rangle_{a_{0}b_{0}c_{0}}\otimes|\Phi\rangle_{a_{1}b_{1}c_{1}}(i=0,1)$
of the composite system composed of the six photons can be rewritten
as
\begin{eqnarray}                              \label{eq2}    % Eq. 9
|\Psi_{1}\rangle \!&=&\! \alpha|RR\rangle_{a_{0}b_{0}}(\beta
|RL\rangle+\gamma|LR\rangle)_{a_{1}b_{1}} |LR\rangle_{c_{0}c_{1}}
\nonumber\\
&&+\alpha(\beta |RL\rangle+\gamma|LR\rangle)_{a_{0}b_{0}}
|RR\rangle_{a_{1}b_{1}}|RL\rangle_{c_{0}c_{1}}\nonumber\\
&&+\alpha^{2}|RR\rangle_{a_{0}b_{0}}|RR\rangle_{a_{1}b_{1}}
|LL\rangle_{c_{0}c_{1}}+(\beta |RL\rangle\nonumber\\
&& +\gamma|LR\rangle)_{a_{0}b_{0}}(\beta
|RL\rangle+\gamma|LR\rangle)_{a_{1}b_{1}}
|RR\rangle_{c_{0}c_{1}}.\;\;\;\;\;\;\;\;
\end{eqnarray}
Charlie lets her two photons $c_{0}$ and $c_{1}$ pass through PCG.
The outcomes will be divided into two groups, the odd-parity one and
the even-parity one.

If the outcome of the PCG by Charlie is an odd-parity one, the state
of the composite system changes from $|\Psi_{1}\rangle$ to the state
(without normalization)
\begin{eqnarray}             \label{eq3}    % Eq. 10
|\Psi_{1o}\rangle \!\!&=&\!\!\alpha(\beta
|RL\rangle+\gamma|LR\rangle)_{a_{0}b_{0}}
|RR\rangle_{a_{1}b_{1}}|RL\rangle_{c_{0}c_{1}}
\nonumber\\
&&\!+\alpha|RR\rangle_{a_{0}b_{0}} (\beta
|RL\rangle+\gamma|LR\rangle)_{a_{1}b_{1}}
|LR\rangle_{c_{0}c_{1}}.\;\;\;\;\;\;\;\;
\end{eqnarray}
Subsequently,  Charlie informs
Alice to measure her photon $a_{1}$ with the basis
$\{|R\rangle,|L\rangle\}$. If Alice gets the outcome
 $|R\rangle_{a_{1}}$ (the single-photon detector $D_{aR}$ is clicked), the state $|\Psi_{1o}\rangle$
 of the composite system becomes
\begin{eqnarray}     % Eq. 11
|\Psi_{1o}^{\prime }\rangle\!\!&=&\!\!\!\nu_1
\big[\beta|RRL\rangle_{a_{0}b_{0}c_{0}}|LR\rangle_{b_{1}c_{1}}
\!+\!(\beta |RLR\rangle\nonumber\\
&&+\gamma|LRR\rangle)_{a_{0}b_{0}c_{0}} |RL\rangle_{b_{1}c_{1}}\big]
\end{eqnarray}
with the probability
$P_{1o}=|\alpha|^{2}(|\gamma|^{2}+2|\beta|^{2})$. Here
$\nu_1=\frac{1}{\sqrt{|\gamma|^{2}+2|\beta|^{2}}}$. After performing
the Hadamard operations $H$ on the photons $b_1$ and $c_1$,  Bob and
Charlie measure their photons $b_1$ and $c_1$ with the basis
$\{\vert R\rangle, \vert L\rangle\}$. When Charlie and Bob obtain an
even-parity one (i.e.,  $|RR\rangle_{b_{1}c_{1}}$ or
$|LL\rangle_{b_{1}c_{1}}$), the three-photon system $a_0b_0c_0$ is
in the state
\begin{eqnarray}                     \label{eq6}     % Eq. 12
|\Phi_{1o}^{+}\rangle &\!=\!\nu_1(\beta |RRL\rangle
\!+\!\beta |RLR\rangle \!+\!\gamma|LRR\rangle)_{a_{0}b_{0}c_{0}}.
\end{eqnarray}
When Charlie and Bob obtain an odd-parity one (i.e.,
$|RL\rangle_{b_{1}c_{1}}$ or $|LR\rangle_{b_{1}c_{1}}$), the
 system  is in the state
\begin{eqnarray}                     \label{eq6'}     % Eq. 13
|\Phi_{1o}^{-}\rangle &\!\!=\!\!\nu_1(-\beta |RRL\rangle
\!+\!\beta |RLR\rangle \!+\!\gamma|LRR\rangle)_{a_{0}b_{0}c_{0}}.
\end{eqnarray}
With a phase-flip operation $\sigma_z=|R\rangle\langle R| -
|L\rangle\langle L|$ on the photon $c_0$, Charlie can transform the
state $|\Phi_{1o}^{-}\rangle$ into the state
$|\Phi_{1o}^{+}\rangle$.

If the outcome of the PCG by Charlie is an even-parity one,  the
composite system collapses from the state $|\Psi_{1}\rangle$ to
(without normalization)
\begin{eqnarray}                   \label{eq7}      % Eq. 14
|\Psi_{1e}\rangle\!\!&=&\!\!-\alpha^{2}|RR\rangle_{a_{0}b_{0}}|RR\rangle_{a_{1}b_{1}}
|LL\rangle_{c_{0}c_{1}}+(\beta |RL\rangle
\nonumber\\
&&\!\!+\gamma|LR\rangle)_{a_{0}b_{0}} (\beta
|RL\rangle+\gamma|LR\rangle)_{a_{1}b_{1}}
|RR\rangle_{c_{0}c_{1}}.\;\;\;\;\;\;\;
\end{eqnarray}
Subsequently, Charlie measures his photons $c_{0}$ and $c_{1}$ with
the basis $\{|R\rangle, |L\rangle\}$. When Charlie gets  the
measurement outcome  $|RR\rangle_{c_{0}c_{1}}$, the state
$|\Psi_{1e}\rangle$ collapses into
\begin{eqnarray}                     \label{eq8}     % Eq. 15
|\Psi_{1e}^{\prime}\rangle =(\beta
|RL\rangle+\gamma|LR\rangle)_{a_{0}b_{0}} (\beta
|RL\rangle+\gamma|LR\rangle)_{a_{1}b_{1}}.
\end{eqnarray}
In this time, Alice and Bob can obtain two pairs of two-photon
systems in  the state $|\Phi_{1e}\rangle=\frac{1}{\sqrt{|\gamma|^{2}+|\beta|^{2}}} (\beta
|RL\rangle+\gamma|LR\rangle)_{\bar{a}\bar{b}}$ with the probability $P_{1e}=(|\gamma|^{2}+|\beta|^{2})^{2}$.

The principle for the second step of our systematic ECP is shown in
Fig.\ref{fig3} (b). In this step, Alice, Bob and Charlie exploit a
set of the three-photon systems in the state $|\Phi_{1o}^{+}\rangle$
and the two-photon systems in the state $|\Phi_{1e}\rangle$ to
finish the task of entanglement concentration for obtaining a subset
of the three-photon systems in a standard W state. To this end,
Alice lets her two photons $a_{0}$ and $\bar{a}$ go through the PCG.
If the photon pair $a_{0}\bar{a}$ is in the odd-parity state, the
five photons system is in the state $|\Psi_{2o}\rangle$ in which
each item has the same parameter, that is,
\begin{eqnarray}                      \label{eq10}     % Eq. 16
|\Psi_{2o}\rangle\!&=&\!\frac{1}{\sqrt{3}}\big[
|RL\rangle_{a_{0}\bar{a}}|R\rangle_{\bar{b}} (|RL\rangle
+|LR\rangle)_{b_{0}c_{0}}\;\;\;\;\;\;\;\;
\nonumber\\
&&\!+|LR\rangle_{a_{1}\bar{a}}|L\rangle_{\bar{b}}
|RR\rangle_{b_{0}c_{0}}\big]
\end{eqnarray}
%It takes place
with the probability
$P'_{1o}=\frac{3|\gamma|^{2}|\beta|^{2}}
{(|\gamma|^{2}+2|\beta|^{2})(|\gamma|^{2}+|\beta|^{2})}$.
Then Alice and Bob perform Hadamard operation and  measure the two photons $\bar{a}$ and $\bar{b}$.
If Alice and Bob obtain
an even-parity outcome, the
three-photon system $a_0b_0c_0$ is in the standard three-photon W
state
\begin{eqnarray}                  \label{eq11}     % Eq. 17
|W^{+}\rangle = \frac{1}{\sqrt{3}}(|RRL\rangle+
|RLR\rangle+|LRR\rangle)_{a_{0}b_{0}c_{0}}.
\end{eqnarray}
If they obtain an odd-parity one, the three-photon system is in the
state $|W^{-}\rangle$
\begin{eqnarray}                  \label{eq11}     % Eq. 18
|W^{-}\rangle = \frac{1}{\sqrt{3}}(|RRL\rangle+
|RLR\rangle-|LRR\rangle)_{a_{0}b_{0}c_{0}}.
\end{eqnarray}
With a phase-flip operation $\sigma_z$ on the photon $a_0$, Alice
can transform the state $|W^{-}\rangle$ into the state
$|W^{+}\rangle$. Therefore, Alice, Bob, and Charlie can get a
three-photon system in a standard $|W^{+}\rangle$ state from the
three-photon systems in a partially entangled state
$|\Phi\rangle_{a_{i}b_{i}c_{i}}$ with the success probability $P_{1}=\xi
P'_{1o}$, where $\xi=\text{min}\{P_{1o}, P_{1e}\}$.
If there are enough three-photon unknown W states and they satisfy
$P_{1o}=P_{1e}$, three participants can obtain the
standard W state with  the success probability is
$P_{1}
=\frac{3|\alpha|^{2}|\beta|^{2}|\gamma|^{2}}{|\gamma|^{2}+|\beta|^{2}}$
in the first round of concentration.

\begin{figure}[htp]
\begin{center}
\includegraphics[width=8 cm,angle=0]{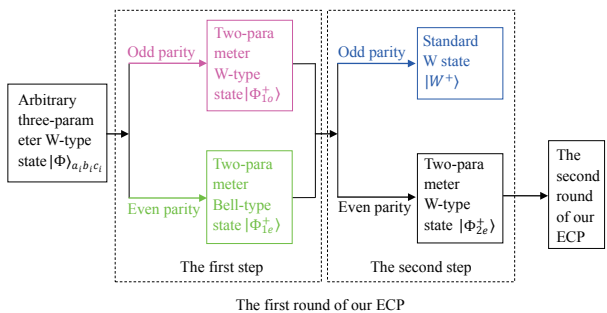}
\caption{ Schematic diagram of the procedure of the first round of
our ECP for three-photon systems in an arbitrary W-type state with
PCG. } \label{fig4}
\end{center}
\end{figure}

If the photon pair $a_{0}\bar{a}$ is in the even-parity one, the
five-photon system is in the state
\begin{eqnarray}            \label{eq12}      % Eq. 19
|\Psi_{2e}\rangle \!&=&\!\nu_2
[\beta^{2}|RR\rangle_{a_{0}\bar{a}}|L\rangle_{\bar{b}}
(|RL\rangle+|LR\rangle)_{b_{0}c_{0}}\nonumber\\
&&\! -\gamma^{2}|LL\rangle_{a_{0}\bar{a}}
|R\rangle_{\bar{b}}|RR\rangle_{b_{0}c_{0}}]
\end{eqnarray}
with the probability $P'_{1e}=\frac{|\gamma|^{4}+2|\beta|^{4}}
{(|\gamma|^{2}+2|\beta|^{2})(|\gamma|^{2}+|\beta|^{2})}$. Here
$\nu_2=\frac{1}{\sqrt{|\gamma|^{4}\!+\!2|\beta|^{4}}}$. Similar to
the above discussion of the measurement results of the odd-parity
case (the state $|\Psi_{2e}$), three participants
obtain the states $|\Phi_{2e}^{+}\rangle$ and
$|\Phi_{2e}^{-}\rangle$ when the outcomes obtained by Alice and
Bob are an odd-parity one and an even-parity one, respectively.
Alice can transform the state $|\Phi_{2e}^{-}\rangle$ into the
state $|\Phi_{2e}^{+}\rangle$ by performing a phase-flip
operation $\sigma_z$ on the photon $a_0$. Here
\begin{eqnarray}              \label{eq13}     % Eq. 20
\begin{split}
|\Phi_{2e}^{+}\rangle &\!=\!\nu_2
(\beta^{2}|RRL\rangle \!+\!\beta^{2}|RLR\rangle +\gamma^{2}|LRR\rangle)_{a_{0}b_{0}c_{0}},\;\;\;\;\\
|\Phi_{2e}^{-}\rangle &\!=\!\nu_2
(\beta^{2}|RRL\rangle\!+\!\beta^{2}|RLR\rangle
-\gamma^{2}|LRR\rangle)_{a_{0}b_{0}c_{0}}.
\end{split}
\end{eqnarray}
It is not difficult to find that the state $|\Phi_{2e}^{+}\rangle$
has the same form as the state $|\Phi_{1o}^{+}\rangle$ shown in
Eq.(\ref{eq6}) but different parameters. We need only replace the
parameters $\beta$  and $\gamma$ in Eq.(\ref{eq6}) with the
parameters $\beta^{2}$  and $\gamma^{2}$, respectively. Obviously,
$|\Phi_{2e}^{+}\rangle$ is the resource for the ECP in the second
round. The above discussion is the first round of our ECP. The
detail for the procedure of the first round of our ECP for
three-photon systems in an arbitrary W-type state with PCG is shown
in Fig.\ref{fig4}.

Now, we continue to discuss the second round of our ECP with the
composite system composed of the six photons $a_{2}b_{2}c_{2}$ and
$a_{3}b_{3}c_{3}$ (different from the previous discussion with six
photons $a_{0}b_{0}c_{0}$ and $a_{1}b_{1}c_{1}$ in the first round)
from a set of the three-photon systems in the state
$|\Phi_{2e}^{+}\rangle$. First of all, we make the six photons
$a_{2}b_{2}c_{2}$ and $a_{3}b_{3}c_{3}$ pass through the device
shown in Fig. \ref{fig3}(a), only substituting $a_{2}, a_{3}, b_{2},
b_{3}, c_{2}, c_{3}$ for $c_{0}, c_{1}, b_{0}, b_{1}, a_{0}, a_{1}$,
respectively. Besides, it's worth noting that two three-photon
systems do not go through the second step of the first round of our
ECP.

\begin{figure}[htp]
\begin{center}
\includegraphics[width=6.4 cm,angle=0]{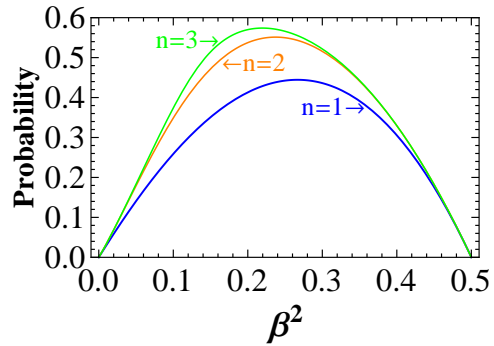}
\caption{ The total success probability of the maximally entangled W
state vs the parameter $\beta^{2}$ with $P_{1o}=P_{1e}$, by
iterating the ECP one time (n=1), two times (n=2), and three times
(n=3), respectively. } \label{fig5}
\end{center}
\end{figure}

Similar to the above discussion in the first round of our ECP, three
parties obtain the standard three-photon W state with the success
probability of
$P_{2o}=\frac{3|\beta|^{4}|\gamma|^{4}}{(|\gamma|^{4}+2|\beta|^{4})^{2}}$.
By iterating the ECP several times, the success probability to get a
maximally entangled W state from the initial partially entangled
state is $P_{no}=\frac{3|\beta|^{2^{n}}|\gamma|^{2^{n}}}
{(2|\beta|^{2^{n}}+|\gamma|^{2^{n}})^{2}}, (n=2, 3,.., )$ in the
n-th round of our ECP, while the probability to obtain the partially
entangled three-photon W states is
$P_{ne}=\frac{2|\beta|^{2^{n+1}}+|\gamma|^{2^{n+1}}}
{(2|\beta|^{2^{n}}+|\gamma|^{2^{n}})^{2}}, (n=2, 3,.., )$.
Therefore, the total success probability to get a maximally
entangled W state from the initial partially entangled state is
\begin{eqnarray}            \label{eq17}     % Eq. 23
P= \xi P'_{1o}+\xi P'_{1e}P_{2o}+\cdot\cdot\cdot+\xi P'_{1e}P_{2e}P_{3e}\cdot\cdot\cdot P_{no}.\;\;
\end{eqnarray}
The total success probability of the maximally entangled W state vs
$\beta^{2}$ with $P_{1o}=P_{1e}$ is shown in Fig. \ref{fig5}. It is
quite clear that the total success probability gradually increases
by iterating the ECP.

\section{Discussion and summary}\label{sec5}

So far, all the procedures in our scheme for the   PCG  are
described in the case that the side leakage rate $\kappa_{s}$ is
negligible. To present our idea  more realistically, $\kappa_{s}$
should be taken into account. In this time, the rules of the input
states changing under the interaction of the photon and the cavity
become
\begin{eqnarray}      % Eq. 24
&& |L,\uparrow\rangle\rightarrow |r(\omega)||L,\uparrow\rangle,
\;\;\;\;\;\;\; |L,\downarrow\rangle\rightarrow
i|r_{0}(\omega)||L,\downarrow\rangle, \;\;\;\;\nonumber\\
&& |R,\uparrow\rangle\rightarrow i|r_{0}(\omega)||R,\uparrow\rangle,
\;\;\;\; |R,\downarrow\rangle\rightarrow
|r(\omega)||R,\downarrow\rangle.\;\;\;\;\;\;\;\;\;\;\;\;
\end{eqnarray}
The fidelity and the efficiency of our PCG are sensitive to
$\kappa_{s}$ as $\kappa_{s}$ influences the amplitudes of the
reflected photon (see Eq.(\ref{RULE})). Here the fidelity of our PCG
are defined as $F=|\langle\psi_{real}|\psi_{ideal}\rangle|^{2}$.
Here, $|\psi_{ideal}\rangle$ and $|\psi_{real}\rangle$ are the final
states in the ideal condition and in the realistic condition,
respectively.  The coupling strength $g/(\kappa_{s}+\kappa)\cong
1.5$  \cite{ReithmaierQD} was reported in $d = 1.5\mu m$ micropillar
microcavities ($Q \sim 8800$), and the coupling strength can be
enhanced to $g/(\kappa_{s}+\kappa)\cong 2.4$ ($Q \sim 40000$)
\cite{YoshieQD} by improving the sample designs, growth, and
fabrication \cite{ReitzensteinQD}. Here the quality factor is
dominated by the side leakage and the cavity loss rate
$(\kappa_{s}/\kappa)$, and $\kappa_{s}/\kappa$ can be reduced by
thinning down the top mirrors, which may decrease the quality
factor, increase $\kappa$, and keep $\kappa_{s}$ nearly unchanged
 \cite{HuQD3}. Using this method, the quality factor $Q \sim 17000$
($g/(\kappa_{s}+\kappa)\cong 1$) was achieved with
$\kappa_{s}/\kappa\sim 0.7$  \cite{HuQD3}.

\begin{figure}
\centering
\includegraphics[width=4.2 cm,angle=0]{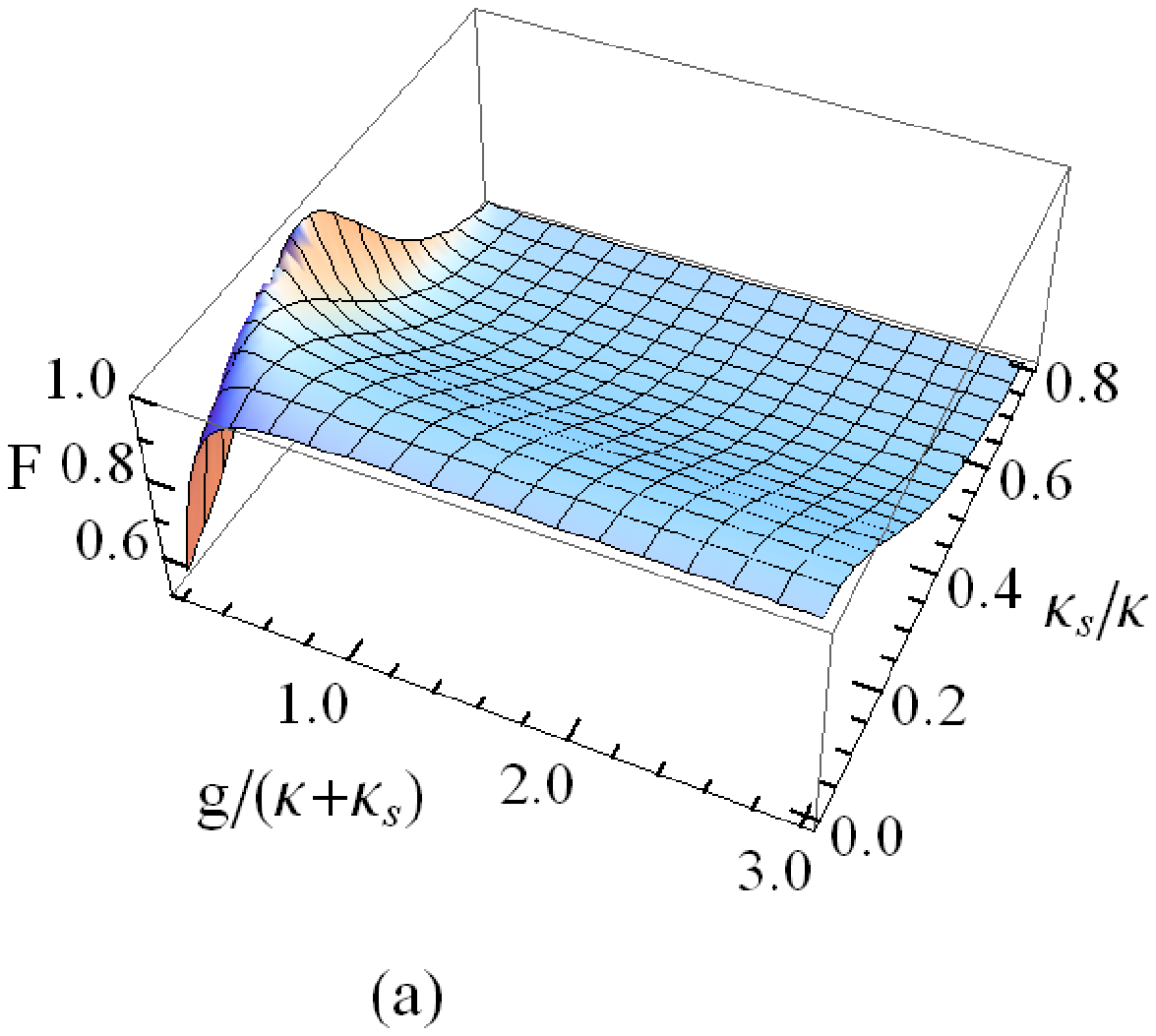}\;\;\;\;%\\
%\bigskip
\includegraphics[width=4.2 cm,angle=0]{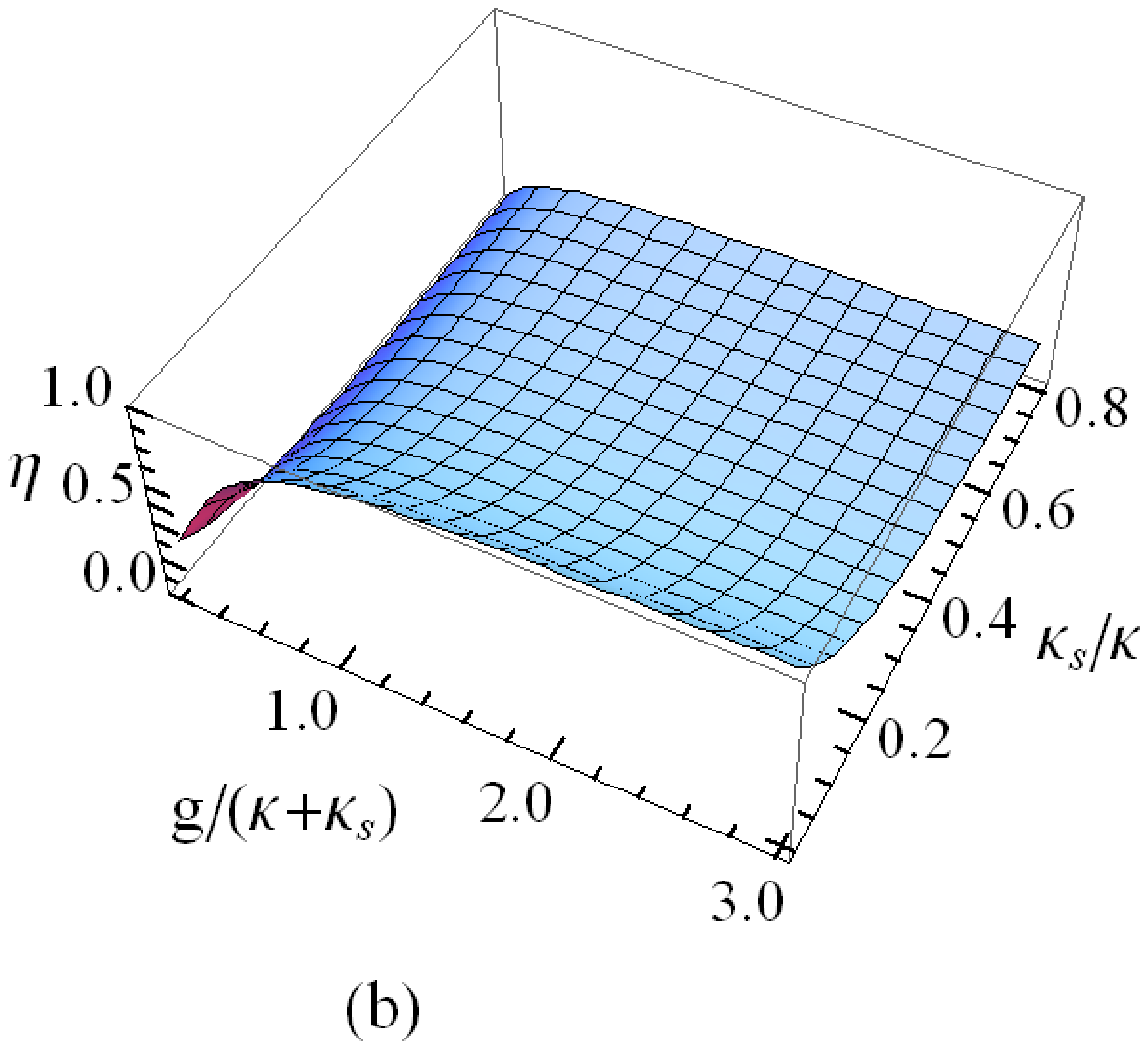}%\\
\caption{ The fidelity (a) and the efficiency (b) of the present
parity-check gate on the polarizations of two photons for our ECP in
the even parity case vs the coupling strength $g/(\kappa
+\kappa_{s})$ and the side leakage rate $\kappa_{s}/\kappa$ with
$\gamma= 0.1\kappa$.}\label{fig6}
\end{figure}

In the condition $g/(\kappa_{s}+\kappa)\cong 2.4$,
$\kappa_{s}/\kappa\sim 0$, and $\gamma\sim 0.1\kappa$, the fidelity
of our PCG in the odd-parity case is robust (almost 1). The fidelity
in the even-parity case and efficiency of our PCG are $F = 100\%$
and $\eta=98.2\%$, respectively. In the case
$g/(\kappa_{s}+\kappa)\cong 1.3$ and $\kappa_{s}/\kappa\sim 0.3$,
the fidelity and efficiency of our PCG are $F = 77.6\%$ and
$\eta=65\% $, respectively.  For the case
$g/(\kappa_{s}+\kappa)\cong 1$ and $\kappa_{s}/\kappa\sim 0.7$, the
fidelity and efficiency of our PCG are $F = 66\%$ and $\eta=45\%$,
respectively. Therefore, the strong coupling and low cavity side
leakage are required in this scheme.  Our scheme is implemented with
a QD-induced phase shift of $\pm\frac{\pi}{2}$, which requires the
frequencies are adjusted to be $\omega-\omega_{c}\approx\kappa/2$
($\omega_{c}=\omega_{X^{-}})$. The fidelity in the even-parity case
and the efficiency of our PCG vary with the coupling strength and
the side leakage rate, and they are shown in Fig. \ref{fig6}(a) and
Fig.\ref{fig6}(b), respectively. From these figures, one can see
that our scheme is feasible in both the strong coupling regime and
the weak coupling regime. $\kappa_{s}$ can be made rather small by
improving the sample growth or the etching process.

Compared with other ECPs for a  W-class state
\cite{Yildiz,wang2,du,ecpaa6,shengpraEC4,ecpaa8,ecpgujosab,ecpaa10,ecpxiayqip},
our ECP has some advantages. First, our ECP  does not require that
two of the three coefficients in the unknown W state are the same
ones  \cite{wang2,du,ecpaa6}. Moreover, it does not require that all
of the coefficients are known for the parties, different from the
ECP for W states in Ref. \cite{shengpraEC4}. Second, our ECP  only
relies on the optical property of the quantum-dot spins inside
one-sided optical microcavities, which is robust in the odd-parity
instance for obtaining the standard W state. As the side leakage and
cavity loss may be difficult to control or reduce for the
electron-spin qubit and photonic qubits in the double-sided
QD-cavity system, our ECP is relatively easy to implement in
experiment. Third, our ECP  requires one of the parties to perform
the local unitary operation and communicate the classical
information with other parties to retain or discard their photons,
which greatly simplifies the complication of classical
communication. Fourth, with nonlinear optical elements, the resource
can be utilized sufficiently and the total success probability of
our ECP  is larger than that in the conventional ECP with linear
optical elements \cite{ecpaa8}, which is caused by preserving the
states that are discarded in the latter. With the iteration of our
ECP process, the success probability $P$ can be increased largely.

In summary, we have proposed a systematic ECP for an arbitrary
unknown less-entangled three-photon W state.  It has some
advantages, compared with others
\cite{Yildiz,wang2,du,ecpaa6,shengpraEC4,ecpaa8,ecpgujosab,ecpaa10,ecpxiayqip}.
First, it has a high efficiency as the parties obtain not only some
partially entangled three-photon systems with two unknown parameters
by picking  up the robust odd-parity instance with PCG, but also
some entangled two-photon systems by keeping an even-parity instance
in the first step of each round of concentration, with which the
parties can obtain a standard three-photon W state. Second, it is a
repeatable one, which can increase the success probability largely.
Third, as the side leakage and cavity loss may be difficult to
control or reduce for the photonic qubits in the double-sided
QD-cavity system, our ECP is relatively easier to be implemented in
experiment than the ECP with a double-sided QD-cavity system
\cite{ecpaa10}.  These advantages maybe make our ECP more useful in
quantum communication network in future.

%
%Moreover, our ECP can obtain the maximally entangled three-photon W
%state, and it can be used to create a private key by resorting to
%quantum key distribution(QKD) protocols \cite{QKD1,QKD2}, even in
%the case with a collective-noise optical-fiber channel \cite{QKD3}.
%With quantum secure direct communication(QSDC) protocols based on
%entangled photon pairs \cite{QSDC1,QSDC2,QSDC3}, multi-user can also
%transmit their secret message directly, without creating a private
%key in advance.

\section*{Acknowledgments}

This work was supported by the National Natural Science Foundation
of China under Grant No. 11174039  and the Open Foundation of State
Key Laboratory of Networking and Switching Technology (Beijing
University of Posts and Telecommunications) under Grant No.
SKLNST-2013-1-13.

\end{document}